\documentclass[10pt,twocolumn,letterpaper]{article}

 \usepackage{cvpr}              
\usepackage{svg}
\usepackage{multirow}
\usepackage[accsupp]{axessibility}

\definecolor{cvprblue}{rgb}{0.21,0.49,0.74}
\usepackage[pagebackref,breaklinks,colorlinks,citecolor=cvprblue]{hyperref}
\usepackage{float}

\def\paperID{9} 
\def\confName{CVPR}
\def\confYear{2024}

\title{SwinFuSR: an image fusion-inspired model for RGB-guided thermal image super-resolution}

\author{Cyprien Arnold\\
Polytechnique Montréal\\
Montréal, Canada\\
{\tt\small cyprien.arnold@polymtl.ca}
\and
Philippe Jouvet\\
CHU Sainte Justine\\
Montréal, Canada\\
{\tt\small philippe.jouvet.med@ssss.gouv.qc.ca}
\and
Lama Seoud\\
Polytechnique Montréal\\
Montréal, Canada\\
{\tt\small lama.seoud@polymtl.ca}
}

\begin{document}
\maketitle
\begin{abstract}
Thermal imaging plays a crucial role in various applications, but the inherent low resolution of commonly available infrared (IR) cameras limits its effectiveness. Conventional super-resolution (SR) methods often struggle with thermal images due to their lack of high-frequency details. Guided SR leverages information from a high-resolution image, typically in the visible spectrum, to enhance the reconstruction of a high-res IR image from the low-res input. Inspired by SwinFusion, we propose SwinFuSR, a guided SR architecture based on Swin transformers. In real world scenarios, however, the guiding modality (e.g. RBG image) may be missing, so we propose a training method that improves the robustness of the model in this case. Our method has few parameters and outperforms state of the art models in terms of Peak Signal to Noise Ratio (PSNR) and Structural SIMilarity (SSIM). In Track 2 of the PBVS 2024 Thermal Image Super-Resolution Challenge, it achieves 3rd place in the PSNR metric. Our code and pretained weights are available at https://github.com/VisionICLab/SwinFuSR.
\end{abstract}    
\section{Introduction}

Improving the quality of digital images is crucial in numerous fields, from mobile photography \cite{10.1007/978-3-031-25066-8_5} and healthcare \cite{peng_da-vsr_2022,gu_medsrgan_2020,qiu_medical_2023} to law enforcement \cite{PAN2023103844}. Super-resolution (SR) has emerged as a promising technique to achieve this goal, allowing the reconstruction of high-resolution (HR) images from their low-resolution (LR) counterparts. In the realm of RGB images, SR has witnessed significant advancements in recent years. Diverse techniques, ranging from traditional methods to deep learning, have been developed to exploit information within LR images and generate realistic detailed HR images.

Infrared (IR) images, capturing the heat emitted by objects, enable night vision and the ability to detect features invisible to the naked eye. The IR modality is used for continuous and contactless monitoring of patients' vital signs in the intensive care units (ICU) \cite{articlePHI,PHI2} and to integrate this information into clinical decision support systems \cite{articleP}. To achieve this, IR acquisitions are often combined with RGB images, and even 3D images \cite{arti}. High definition IR sensors with spatial resolution of up to 1,024 × 768 pixels are commercially available, but can cost tens of thousands of dollars. Hence, lower resolution IR sensors tend to be used instead in ICU rooms. 

Thermal image super-resolution (TISR) tackles this challenge by increasing image resolution and revealing details obscured in the LR image. This topic is increasingly studied because of its many applications \cite{huang_infrared_2022} including in medical science \cite{article,senalp_new_2023}, agricultural management \cite{CAO2020105331,s20236732} or even space studies \cite{Yamaguchi_2020,2012ApJ...749L..20H}. Several challenges remain in fully realizing the potential of IR super-resolution. One key challenge lies in the inherent differences between IR and RGB images. IR images exhibit higher noise and poorer texture information \cite{huang_infrared_2022}, making HR reconstruction more complex.

Guided thermal image super resolution (GTISR) presents itself as a particularly promising approach for IR image reconstruction. By relying on an HR reference image as input, such as a corresponding visible spectrum image, guided SR can improve the accuracy and consistency of the reconstruction. In effect, HR RGB images are cost-effective to obtain and have higher frequencies than IR images. To encourage researchers to innovate in this little-explored field, the 19th IEEE Workshop on Perception Beyond the Visible Spectrum introduced a challenge track \cite{rivadeneira_thermal_2022} in 2023 to generate x8 super-resolution thermal images by using visible HR images as guidance. Candidates are ranked according to Peak Signal to Noise Ratio (PSNR) and Structural Similarity Index Measure (SSIM) between images produced from the test set and the (non-public) ground truth HR IR images. 

In this paper, we draw inspiration from multimodal image fusion based on Swin Transformers to propose SwinFuSR, a novel method for RGB-guided thermal image super-resolution. Our contributions are two-fold:
\begin{itemize}
\item a lightweight transformer-based model that outperforms other state of art GTISR methods.
\item a modified training strategy that improves the robustness of GTISR in the absence of the guiding modality.
\end{itemize}

\section{Related Works}

\subsection{Visible image super-resolution}
The first approaches to super resolution employed so-called ``traditional'' approaches \cite{huang_infrared_2022}. These methods either focused on the frequency domain, trying to model the relationship between the HR and LR images using mathematical models \cite{aizawa1991acquisition,rhee1999discrete,shen2007map}, or used dictionaries methods to map LR patches to HR patches \cite{yang2012coupled,wang2012semi,zhang2011efficient}.

2015 saw the emergence of deep learning-based methods using convolutional neural networks (CNNs) such as SRCNN \cite{dong2015image}, FSRCNN \cite{dong2016accelerating} and ESPCN \cite{shi2016real}, which introduced subpixel convolutional layers, a new upsampling operation. The advent of residual networks \cite{he2016deep} (to solve the vanishing gradient problem in particular) led to new architectures like VDSR \cite{kim2016accurate}, RED \cite{mao2016image} and EDSR \cite{lim2017enhanced}, the latter proposing a new residual connection and winning the NTIRE2017 Super-Resolution Challenge \cite{8014883}.

In 2017, SRGAN \cite{ledig2017photo} achieved remarkable results by applying a generative adversarial network (GAN) to the SR task. One year later appeared ESRGAN\cite{wang2018esrgan}, an enhanced version of SRGAN.

Since transformers \cite{vaswani2017attention} have been adapted to the field of computer vision with the Vision Transformer ViT \cite{dosovitskiy2020image}, the Swin transformer \cite{liu2021swin} resolved the computational complexity problem of ViT by using shifted windows. With this mechanism, SwinIR \cite{liang2021swinir} applied the Swin architecture to the image reconstruction task and outperformed the best existing architectures. SwinIR's main strength lies in its Residual Swin Transformer Blocks, which extract highly relevant features. More recently, the HAT  \cite{chen2023activating} and SwinFIR \cite{zhang_swinfir_2023} architectures have proposed improvements to SwinIR and represent the current state of the art in SR.

\subsection{Thermal image super-resolution}
Compared to RGB images, IR images are single channel, have low gradients and ``overlapping information between high and low frequencies'' \cite{huang_infrared_2022}. To manage these characteristics, specific architectures have been proposed for IR images.

Before deep learning, frequency domain-based solutions like \cite{wang2009analysis} or dictionary-based methods \cite{deng2014infrared} were proposed. Then, inspired by the methods used in the visible spectrum, \cite{fan2021infrared} and \cite{zou2021super} exploited CNNs and residual networks. Other architectures have come up with the idea of using visible information (more abundant data) to reconstruct the IR image. For example, \cite{yang2020deep} used visible information in the loss function, while PSRGAN \cite{huang_infrared_2021} used a GAN framework and transfer learning from RGB images to train their SR algorithm.

More recently, approaches using transformers have appeared, namely DASR \cite{liang_dasr_2023} that exploits spatial and channel attention. In the same spirit, \cite{zhang_thermal_2023} dynamically reweights the output of attention and non-attention branches to improve the resolution and restore high-frequency details, offering a lightweight structure suitable for edge device deployment.

\subsection{Guided thermal super-resolution} 
Unlike the methods presented above, guided methods take two paired images as input: an LR thermal image (or target image) and a higher-resolution guide image to help with the SR task. One of the first GTISR works was introduced in \cite{Zou2020AnII}; it used a dual-path residual network to merge features from the visible and IR domains. More recently, CoRefusion \cite{kasliwal_corefusion_2023} has been proposed. Its architecture is composed of two U-Nets \cite{ronneberger2015unet} with residual connections to fuse both modalities. A contrastive term is added to the loss function and yields improved performance. 

CoreFusion was part of the first GTISR track in the 2023 PBVS competition \cite{rivadeneira_thermal_2022}. However, the winner of that competition was GuidedSR \cite{rivadeneira_thermal_2022}; this latter approach concatenates RGB and IR features from the shallow feature extraction layers and uses Non-linear Activation Free (NAF) blocks \cite{chen_simple_2022} to fuse RGB and IR information.

More recently, the authors of \cite{suarez_enhancement_2024} described several SR guided methods applied to thermal images and how transforming the RGB guide image into a "thermal-like image" improves performance. They show that this substitution boosts performance by a few percentage points in different super-resolution guided architectures.

\subsection{Multimodal image fusion}
Multimodal image fusion aims at combining relevant information from images acquired with different sensors into a single image. DL-based fusion methods can be divided into three categories: early fusion, late fusion and hybrid fusion\cite{bayoudh_survey_2022}. The first one merges features before task related layers, the second one uses task related layers on each modality before aggregating the information, while the last one combines the first two approaches.

SwinFusion \cite{ma_swinfusion_2022} proposed to fuse images from two modalities using Attention-guided Cross-domain Fusion (ACF). Inspired by the Swin Transform block, this model merges information from the two modality branches via alternating modules of ``self-attention-based intra-domain fusion'' and ``cross-attention-based inter-domain fusion'' units.

The work in \cite{liu_target-aware_2022} proposed a Target-aware Dual Adversarial Learning for object detection. The idea is to exploit structural information in the IR image and textural details from the visible image to improve object detection. This is made possible by means of a generator and two discriminators that seek to retain relevant information from the two modalities.

\subsection{Robustness to missing imaging modality}
GTISR is subject to degraded performance when one of the inputs, e.g. the guiding RGB image, is missing at inference time. Little work in the literature addresses this issue directly for the thermal image SR task, but some studies have examined it in other application areas.

In \cite{woo2023good}, the authors evaluated the impact of the type of architecture, data augmentation and image fusion technique on action recognition performance in the case of a missing modality. They concluded that transformer-based fusion is more robust in this situation than feature summation or concatenation. Meanwhile, \cite{ma_smil_nodate} studied the impact of a missing modality (text, audio, or image) in training or testing a GAN or autoencoder model. They proposed a Bayesian meta-learning framework to better manage missing modalities. 
\section{Method}
\begin{figure*}[ht]
  \centering
  
   \includegraphics[width=17cm]{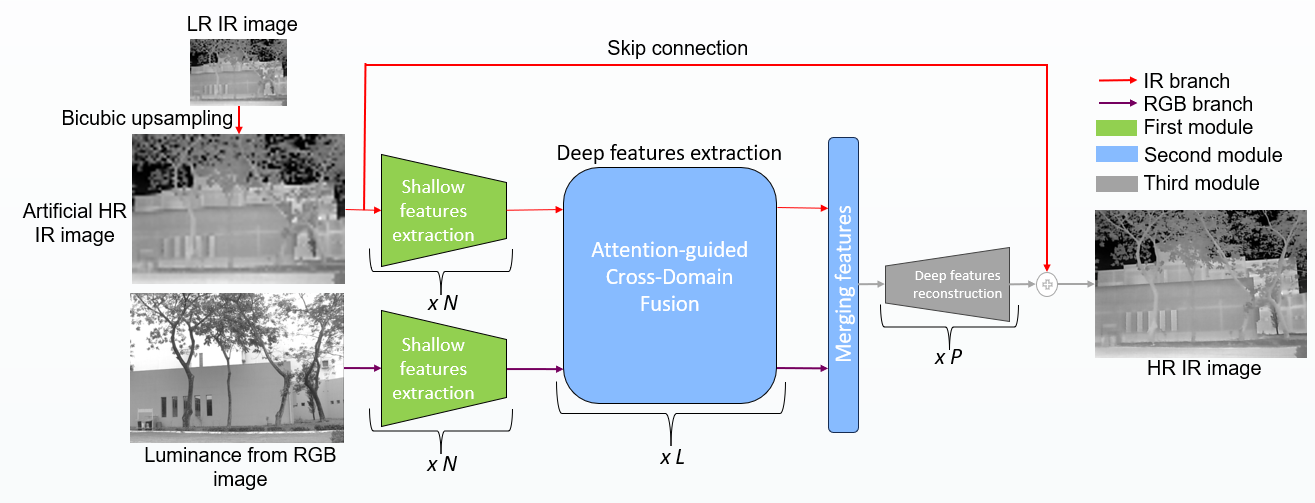}
   \caption{Architecture of the proposed SwinFuSR model.}
   \label{fig:architecture}
\end{figure*}

In this paper, we propose a novel architecture, named SwinFuSR, as a contender for the PBVS 2024 TISR Track 2 challenge. The aim of this competition is to obtain a high-resolution (x8) infrared image from a low-resolution IR image and a medium-resolution RGB image.

\subsection{Proposed architecture}
As in many other super resolution transformer architectures \cite{liang2021swinir,rivadeneira_thermal_2022,chen2023activating}, our own, illustrated in \cref{fig:architecture}, is composed of three modules.

The first module extracts shallow features using convolutional layers followed by $N$ Swin Transformer (STL) layers.
The second module focuses on deep feature extraction. Its role is to extract characteristics that are useful to reconstruct the image by combining IR and RGB features. $L$ Attention-guided Cross-domain Fusion (ACF) blocks are used to extract useful information from RGB and IR features. Then, concatenation and convolution are performed to merge the two branches.
The third module carries out deep feature reconstruction. It is composed of $P$ Swin Transformer layers to refine the merged features and three convolution layers to return to image space. 

In the first two modules, the architecture is divided into two branches, similarly to SwinFusion \cite{ma_swinfusion_2022}: one dedicated to the RGB image and the other to the IR image. A bicubic interpolation is performed on the IR image so that its dimensions (height ($h$) and width $w$) match those of its paired RGB image. Inspired by \cite{liang2021swinir,zou2021super,wang2018esrgan}, a skip connection from the interpolated IR image to the reconstructed image is introduced for faster convergence and better performance. This gives the network an initial solution to improve upon.

\subsection{Loss function}
As a loss function, we use a combination of two differentiable pixel losses commonly used to measure the similarity between two images:
\begin{itemize}
    \item An $L_1$ loss (or MAE) allows for relatively stable convergence and avoids gradient explosion \cite{yu_review_2023}:
\begin{equation}
  L_{1} = \frac{1}{n} \sum^{n}_{i=1} \lvert y_{i} - \hat{y_{i}}\rvert
  \label{eq:L1 loss}
\end{equation} with $n$ the number of pixels, $y_{i}$ the value of the $i^{th}$ pixel in the ground-truth (GT) image and $\hat{y_{i}}$ the value of the $i^{th}$ pixel in the reconstruction.\\
    \item An $L_2$ loss (or MSE) is more sensitive to  
higher reconstruction errors but can make the reconstruction smoother at the expense of valuable high-frequency details:
\begin{equation}
  L_{2} = \frac{1}{n} \sum^{n}_{i=1} ( y_{i} - \hat{y_{i}})^2
  \label{eq:L2 loss}
\end{equation}
\end{itemize}

The lower these two metrics are, the closer the reconstruction is to the GT.\\

We use the loss strategy proposed by GuidedSR, the winning solution in the 2023 PBVS challenge, described in \cite{rivadeneira_thermal_2022}.
\begin{equation}
    Loss =
   \begin{cases}
      L_{1} & \text{for the first $T$ epochs}\\
      L_{2} & \text{after}
    \end{cases}       
  \label{eq:Loss strategy}
\end{equation}
This strategy allows us to obtain good convergence properties with an $L_1$ loss, then refine the optimization with an $L_2$ loss.

\subsection{Training strategy}
Specific training strategies can help build missing modality robustness into the model. 
The literature proposes two main ways to handle this. The first one is to remove the entire portion of the network dealing with the missing information; in that case, the modalities must be processed independently as in CoRefusion\cite{kasliwal_corefusion_2023}. The second and simpler method is to arbitrarily set the corresponding input values to the network to a fixed value such as zero.

To reduce performance degradation in the case of a missing modality, we propose a new model training regime that consists in randomly removing the training RGB images. More formally, at each training iteration, the input $I$ to the network is given by:
\begin{equation}
    I =
   \begin{cases}
      (I^{IR}_{h,w},I^{RGB}_{h,w}) &  \text{\space if $p < p_{th}$} \\
      (I^{IR}_{h,w},O_{h,w}) &  \text{otherwise}
    \end{cases}
  \label{eq:Missing mod}
\end{equation}
with $I^{RGB}_{h,w}$ the RGB image, $I^{IR}_{h,w}$ its corresponding IR image after bicubic interpolation, $O_{h,w}$ an all-zero image, $p$ a random probability following a uniform distribution $\mathcal{U}(0, 1)$ and $p_{th}$ a fixed threshold between 0 and 1.

\section{Experiments}
\subsection{Implementation details}
To train our model, we used the dataset provided for the second track of the PBVS 2024 TISR Challenge. It is composed of 700 training samples and 200 validation samples, each sample being a 640x448 IR image, along with its downsampled version by a factor of 8 and its paired 640x448 RGB image. The 100 testing samples are provided without the HR ground truths. These registered images were acquired by Balser (for RGB) and TAU2 (for IR) cameras and represent images of outdoor urban scenes. We evaluated our model's performance on the training and validation sets using the PSNR and SSIM metrics.

Following common practice for training transformer architectures \cite{ma_swinfusion_2022,zhang_swinfir_2023,liang2021swinir,liang_dasr_2023}, we used patches rather than the entire image as input. The patch size used was 128x128 and batch size was 16. The input patches were augmented with random horizontal and vertical flips and random rotations. Pixel values were normalized between 0 and 1.

The number of heads, the window size and the embedding dimensions were set to 6, 9 and 60 respectively. We set the network module depths to $N=2$, $L=3$ and $P=3$, according to the study detailed in \Cref{modules} below.

For the training, the learning rate was set to $4\times10^{-4}$ until $T=3300$, then reduced to  $1\times10^{-4}$ for the remainder. We used the Adam optimizer. The run lasted 72 hours (4300 epochs) on two Tesla V100 GPUs with 32.0 GB of VRAM each. 

\subsection{Ablation study}
\subsubsection{Effect of the number of modules}\label{modules}
\begin{figure}[!ht]
    \centering
    \begin{subfigure}[b]{0.45\textwidth}
       \includegraphics[width=3.0in]{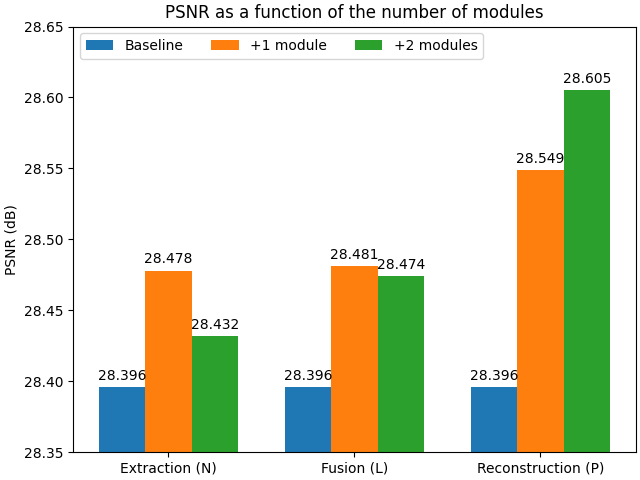}
    \end{subfigure}
    \begin{subfigure}[b]{0.45\textwidth}
      \includegraphics[width=3.0in]{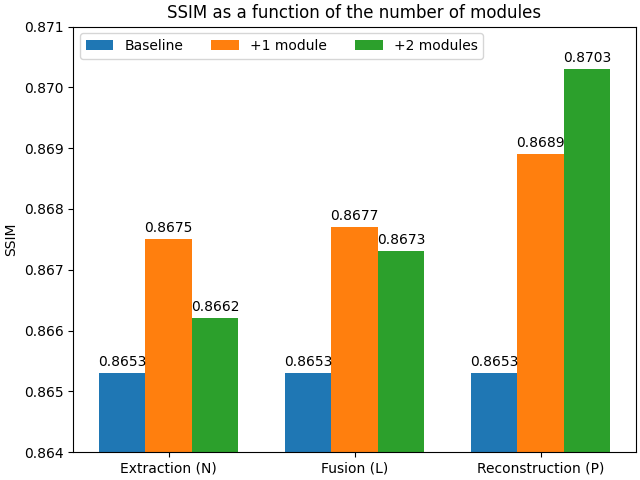}
    \end{subfigure}
    \caption{Effect of module depth on overall performance.}
    \label{fig:ablation}
\end{figure}

To study the effect of the number of STL blocks ($N$), ACF blocks ($L$) and STL blocks ($P$) in the extraction, fusion and reconstruction modules respectively, we set as a baseline $N=1, L=2, P=1$ as in the original SwinFusion paper \cite{ma_swinfusion_2022}. Then, we increased for each module separately these values by 1 and by 2 and observed the effect on performance (PSNR and SSIM) (see \Cref{fig:ablation}).

We can see that the increase in performance is most visible in the reconstruction module, suggesting that the latter is the network bottleneck. Increasing the number of modules in the extraction and fusion modules by 1 each also improves performance, but to a lesser degree. Based on these results, we set the numbers of modules to N = 2, L = 3 and P = 3 for the experiments in \Cref{last} below. For the remaining experiments, we set them to N = 1, L = 2 and P = 1 to limit required resources.

\subsubsection{Effects of skip connection}
\begin{figure}[ht]
    \centering
    \begin{subfigure}[b]{0.45\textwidth}
\includegraphics[width=3.0in,height=2.0in]{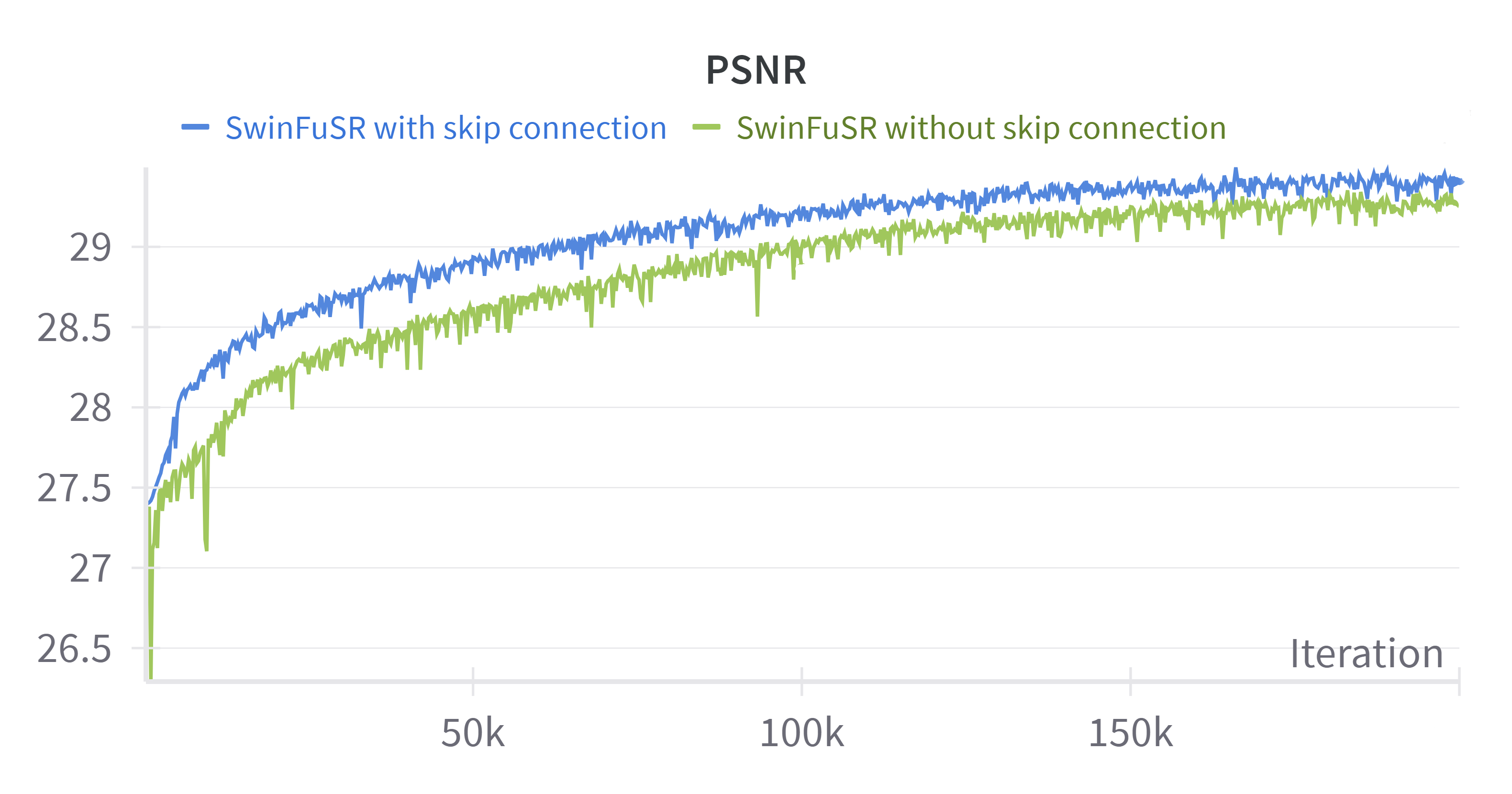}
    \end{subfigure}
    \begin{subfigure}[b]{0.45\textwidth}
\includegraphics[width=3.0in,height=2.0in]{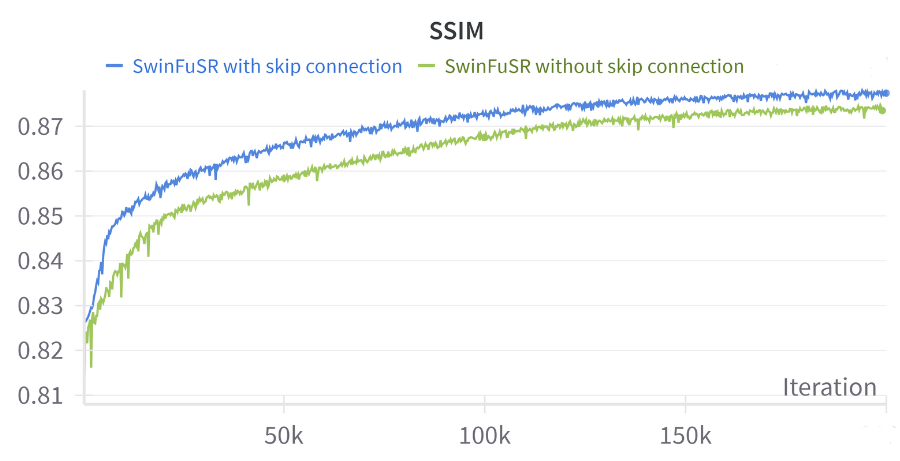}
    \end{subfigure}
    \caption{Performance with (blue) and without skip connection (green).}
    \label{fig:skip connection}
\end{figure}

In SR, it is common to use skip connections between an artificial upsampling or early feature extraction layer and the end of the network. We trained our model with and without the skip connection in our SwinFuSR model (\Cref{fig:architecture}). \Cref{fig:skip connection} shows the difference in performance.
\\
The results demonstrate that using a skip connection improves the convergence speed of the model and improves final performance by 0.3\%. It is important to note that this performance enhancement does not come at the cost of additional parameters in the model.

\section{Results and discussion}

\subsection{RGB guided thermal image super-resolution} \label{last}

For a fair comparison between our solution and the existing methods GuidedSR and CoRefusion, we  retrained the latter two models on the PBVS24 Track 2 dataset, using the same training setup as originally described in their respective papers \cite{kasliwal_corefusion_2023,rivadeneira_thermal_2022} (no pre-trained weights were available). Quantitative results are provided in \Cref{tab:Quantitave result}.

\begin{table}[ht]
\begin{tabular}{ |p{2.5cm}||p{1cm}|p{1cm}||p{1.8cm}|  }
 \hline
 Method& PSNR& SSIM& \#parameters\\
 \hline
 Bicubic   &  25.17    & 0.774& $\varnothing$\\
 CoReFusion&   27.27  & 0.835 &46.31M \\
 GuidedSR &27.22 & 0.834 & 116.35M\\
 SwinFuSR (ours)   & \textbf{28.96} & \textbf{0.878}&\textbf{3.30M}  \\

 \hline
 \end{tabular}
 \caption{PSNR and SSIM on validation set.}
 \label{tab:Quantitave result}
 \end{table}
 
\begin{figure*}[ht]
  \centering
  \begin{tabular}[c]{cccc}
    \multirow{2}{*}[14pt]{
    \begin{subfigure}{0.30\textwidth}
      \includegraphics[height=3cm, width=\textwidth]{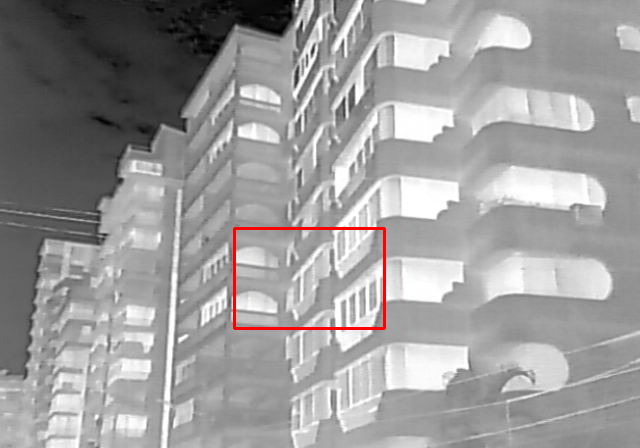}
      \caption{HR}
    \end{subfigure}
}&
   \begin{subfigure}[c]{0.2\textwidth}
      \includegraphics[height=1.5cm, width=\textwidth]{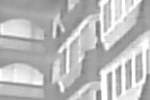}
       \caption{HR}
      \label{fig:hr_scene0}
    \end{subfigure}&
    \begin{subfigure}[c]{0.2\textwidth}
      \includegraphics[height=1.5cm, width=\textwidth]{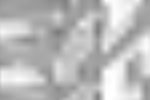}
      \caption{LR /8}
    \end{subfigure}&
    \begin{subfigure}[c]{0.2\textwidth}
      \includegraphics[height=1.5cm, width=\textwidth]{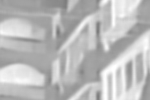}
      \caption{GuidedSR}
    \end{subfigure}\\
    &
        \begin{subfigure}[c]{0.2\textwidth}
      \includegraphics[height=1.5cm, width=\textwidth]{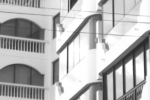}
      \caption{Luminance}
    \end{subfigure}&
    \begin{subfigure}[c]{0.2\textwidth}
      \includegraphics[height=1.5cm, width=\textwidth]{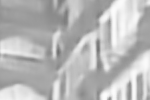}
      \caption{CoReFusion}
    \end{subfigure}&
    \begin{subfigure}[c]{0.2\textwidth}
      \includegraphics[height=1.5cm, width=\textwidth]{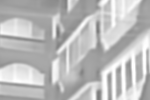}
       \caption{SwinFuSR (ours)}
    \end{subfigure}
  \end{tabular}    
  \caption{GTISR on image 292\_01\_D4 from PBVS 2024 Track-  dataset.}
  \label{fig:GSR}
\end{figure*}

\begin{figure*}[h]
  \centering
  \begin{tabular}[c]{ccccc}
    \begin{subfigure}[c]{0.19\textwidth}
      \includegraphics[height=1.2\linewidth, width=0.9\linewidth]{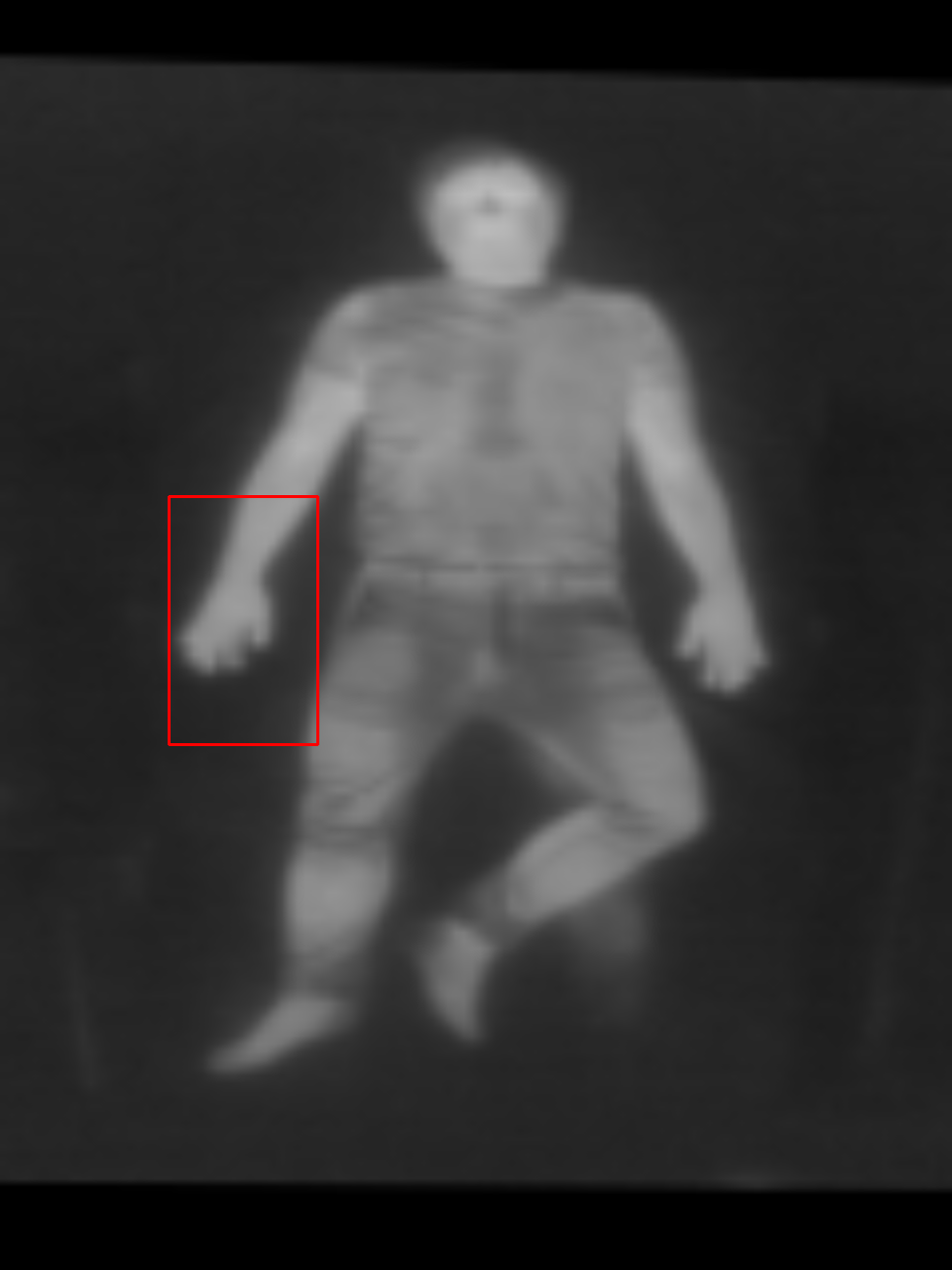}
      \caption{Original LR image}
      \label{fig:ceoa}
    \end{subfigure}&
   \begin{subfigure}[c]{0.19\textwidth}
      \includegraphics[height=1.2\linewidth, width=0.9\linewidth]{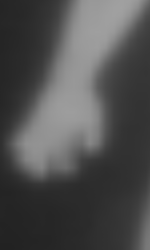}
       \caption{LR}
    \end{subfigure}&
    \begin{subfigure}[c]{0.19\textwidth}
      \includegraphics[height=1.2\linewidth, width=00.9\linewidth]{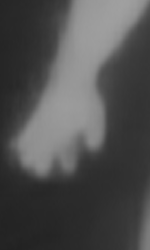}
      \caption{GuidedSR}
    \end{subfigure}&
     \begin{subfigure}[c]{0.19\textwidth}
      \includegraphics[height=1.2\linewidth, width=0.9\linewidth]{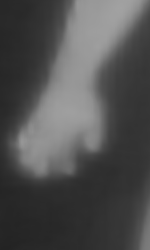}
      \caption{CoReFusion}
    \end{subfigure}&
    \begin{subfigure}[c]{0.19\textwidth}
      \includegraphics[height=1.2\linewidth, width=0.9\linewidth]{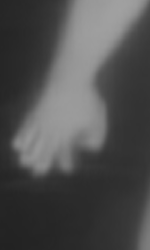}
      \caption{SwinFuSR (ours)}
    \end{subfigure}
    
    \\
  \end{tabular}    
  \caption{GTISR on sample image from SLP dataset \cite{Liu_2019}.}
  \label{fig:GSR SLP}
\end{figure*}

 \Cref{fig:GSR} provides some qualitative results for guided SR on an image from the PBVS2024 challenge dataset. We can notice that SwinFuSR offers the closest output to the GT and seems clearer than the other 2 reconstructions.\\
 To test our solution on different kinds of images and to verify generalization capabilities, we applied guided SR on images from the Simultaneously-collected multimodal Lying Pose (SLP) dataset \cite{Liu_2019}. This dataset is composed of low-resolution (120x160) infrared and RGB image pairs of adult subjects lying down in a hospital bed. \Cref{fig:GSR SLP} shows the results of the x8 guided SR of an image from this dataset. Unfortunately no GT IR images of higher resolution are provided in SLP. Thus, we used the available images as is but could not compare the SR results to reference HR images.
 Qualitatively, all three SR solutions enhance the very low-quality original image. Nevertheless, the details of the hand generated by SwinFuSR seem to be the most accurate, even if the shape of the hand seems unrealistic.
 
\begin{figure*}[!htb]
  \centering
  \begin{tabular}[c]{cccc}
    \multirow{2}{*}[14pt]{
    \begin{subfigure}{0.30\textwidth}
      \includegraphics[height=3cm, width=\textwidth]{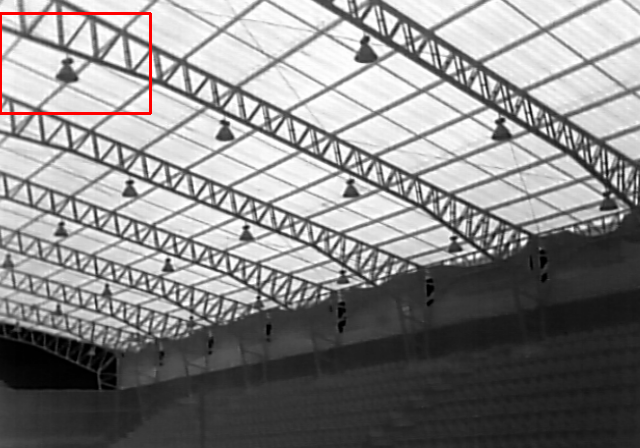}
      \caption{HR}
    \end{subfigure}
}&
   \begin{subfigure}[c]{0.2\textwidth}
      \includegraphics[height=1.5cm, width=\textwidth]{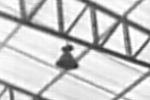}
       \caption{HR}
    \end{subfigure}&
    \begin{subfigure}[c]{0.2\textwidth}
      \includegraphics[height=1.5cm, width=\textwidth]{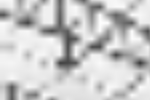}
      \caption{LR /8}
    \end{subfigure}&
    \begin{subfigure}[c]{0.2\textwidth}
      \includegraphics[height=1.5cm, width=\textwidth]{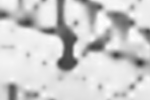}
      \caption{SwinFuSR base}
      \label{fig:swinfu}
    \end{subfigure}
    \\
    &
    \begin{subfigure}[c]{0.2\textwidth}
      \includegraphics[height=1.5cm, width=\textwidth]{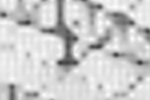}
      \caption{GuidedSR}
    \end{subfigure}
        &
    \begin{subfigure}[c]{0.2\textwidth}
      \includegraphics[height=1.5cm, width=\textwidth]{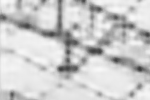}
      \caption{CoReFusion}
    \end{subfigure}&
    \begin{subfigure}[c]{0.2\textwidth}
      \includegraphics[height=1.5cm, width=\textwidth]{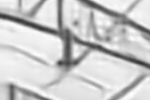}
       \caption{SwinFuSR $p_{th}=0.5$}
      \label{fig:swinfu05}
    \end{subfigure}
  \end{tabular}    
  \caption{Unguided super resolution on image 044\_02\_D1 from PBVS 2024 Track 2 dataset.}
  \label{fig:USR}
\end{figure*}
\begin{figure*}[!htb]
    \centering
    \begin{subfigure}[b]{0.48\textwidth}
      \includegraphics[width=2.8in]{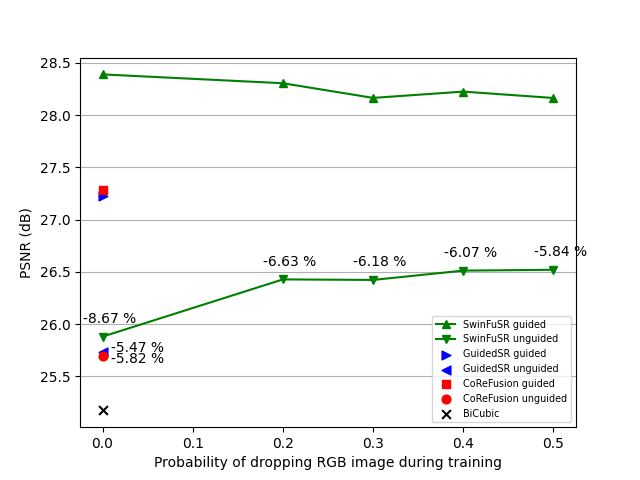}
    \end{subfigure}
    \begin{subfigure}[b]{0.48\textwidth}
      \includegraphics[width=2.8in]{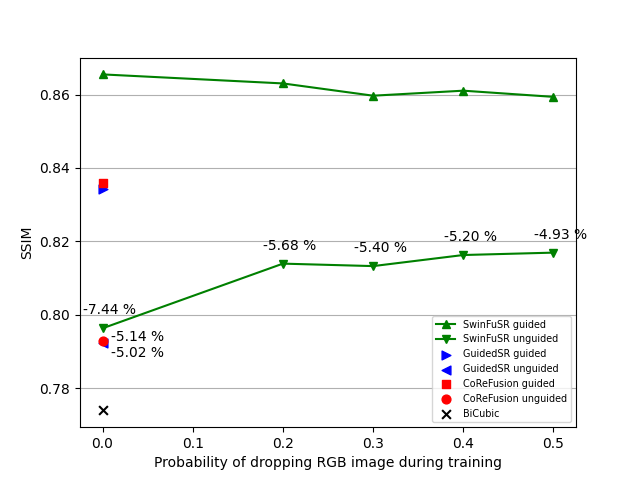}
    \end{subfigure}
    \caption{Effect of training parameter $p_{th}$ on performance with (SwinFuSR guided) and without (SwinFuSR unguided) RGB input images at inference on the PBVS24 validation set.}
    \label{fig:ablation proba}
\end{figure*}
\subsection{Robustness to missing RGB modality}

To evaluate our proposed training regime to improve robustness to missing RGB input, we trained our network five times, each with a different probability threshold $p_{th}$. \Cref{fig:ablation proba} illustrates the performance with and without RGB guide images at inference on the PBVS24 validation set. $p_{th}=0$ means that during the training, all RGB guide images were used during the training.

First of all, we note that GuidedSR and CoReFusion have a smaller drop in performance than SwinFuSR when removing the RGB guide images. We can explain this by the fact that their baseline performance in guided SR is much lower than SwinFuSR.

Second, we see that increasing $p_{th}$ substantially improves the performance of SwinFuSR when no guide image is used for inference (from -8.67\% to -6.63\% for PSNR and from -7.44\% to -5.68\% for SSIM) when $p_{th}$ goes from 0 to 0.2, while only slightly reducing performance for guided SR in terms of both metrics. 
This result suggests that dropping RGB images during training with a certain probability enables a trade-off between maintaining good performance in guided SR and improving results in the absence of  guide images.\\
\Cref{fig:USR} confirms visually that this training strategy can increase performance in unguided SR. Indeed, \Cref{fig:swinfu05} is much clearer than \Cref{fig:swinfu}.


\subsection{Discussion}
Our model is much smaller in terms of parameters than the two competing methods (CoReFusion and GuidedSR), but is slower at inference (1.3s to go from 80x56 to 640x448 running on a PC equipped with an RTX 3080 GPU and 12 GB of VRAM). This limitation restricts the use of SwinFuSR for real-time inference. Moreover, model selection (varying the number of blocks) is costly in terms of VRAM usage, and required us to run those experiments on a GPU cluster. These two drawbacks are probably due to the high proportion of transformers in the network, which are known to be particularly resource-hungry.\\
Another aspect to consider in order to efficiently use the proposed architecture on other datasets is the fact that the IR and RGB images must be registered. For this purpose, several algorithms are available, such as the one proposed in \cite{genser_camera_2020} or Elastix \cite{klein_elastix_2010}, the method used in the PBVS competition. In future work, we will study the robustness of the proposed model to IR-RGB registration errors. 

\section{Conclusion}
This article proposes a new method for RGB guided thermal image super resolution. Our solution, named SwinFuSR, was submitted to Track 2 of the PBVS 2024 Thermal Image Super-Resolution Challenge and achieved better qualitative and quantitative results than other state-of-the-art architectures.  
We also present a novel training strategy that improves robustness to missing guide images at inference time. By randomly dropping a portion of the RGB images during training, the model's performance in unguided SR 
improves significantly
compared to the guided SR baseline.\\
In future work, we will explore how to make better use of the RGB image data, for instance by generating pseudo-IR images. In addition, we will examine how super resolution can improve the performance of related tasks such as estimating in-bed human pose.

\section{Acknowledgments}

We thank Philippe Debanné for his valuable help in editing this paper. The project was supported by L. Seoud's NSERC Discovery grant. This research was enabled in part by support provided by the Digital Research Alliance of Canada (alliancecan.ca).

{
    \small
    \bibliographystyle{ieeenat_fullname}
    \bibliography{main}
}


\end{document}